\documentclass[aps,prl,twocolumn, superscriptaddress,showpacs,showkeys]{revtex4}

\usepackage[dvips]{graphicx,color}
\newcommand{\urusi}{URu$_2$Si$_2$}

\newcommand{\pgnfigure}[2]{\begin{figure}\includegraphics[width=6.5cm,clip=true]
{#1.eps}\caption{\label{#1}#2}\end{figure}}

\begin{document}

\title{Parasitic small-moment-antiferromagnetism and
non-linear coupling of hidden order and antiferromagnetism in
URu$_2$Si$_2$ observed by Larmor diffraction}

\author{P. G. Niklowitz}
\affiliation{Physik Department E21, Technische Universit\"at
M\"unchen, 85748 Garching, Germany} \affiliation{Department of
Physics, Royal Holloway, University of London, Egham TW20 0EX, UK}

\author{C. Pfleiderer}
\affiliation{Physik Department E21, Technische Universit\"at
M\"unchen, 85748 Garching, Germany}

\author{T. Keller}
\affiliation{ZWE FRM II, Technische Universit\"at M\"unchen, 85748
Garching, Germany} \affiliation{Max-Planck-Institut f\"ur
Festk\"orperforschung, Heisenbergstrasse 1, 70569 Stuttgart,
Germany}

\author{M. Vojta}
\affiliation{Institute for Theoretical Physics Universit\"at zu
K\"oln, Z\"ulpicher Strasse 77, 50937 K\"oln, Germany}

\author{Y.-K. Huang} \affiliation{Van der Waals-Zeeman Institute, University of Amsterdam, 1018XE Amsterdam, The
Netherlands}

\author{J. A. Mydosh} \affiliation{Kamerlingh
Onnes Laboratory, Leiden University, 2300RA Leiden, The
Netherlands}

\begin{abstract}
We report simultaneous measurements of the distribution of lattice
constants and the antiferromagnetic moment in high-purity
URu$_2$Si$_2$, using both Larmor and conventional neutron
diffraction, as a function of temperature and pressure up to
18\,kbar. We establish that the tiny moment in the hidden order
(HO) state is purely parasitic and quantitatively originates from
the distribution of lattice constants. Moreover, the HO and
large-moment antiferromagnetism (LMAF) at high pressure are
separated by a line of first-order phase transitions, which ends
in a bicritical point. Thus the HO and LMAF are coupled
non-linearly and must have different symmetry, as expected of the
HO being, e.g., incommensurate orbital currents, helicity order,
or multipolar order.
\end{abstract}


\maketitle


In recent years hydrostatic pressure has become widely used in the
search for new forms of electronic order, because it is believed
to represent a controlled and clean tuning technique. Novel states
discovered in high-pressure studies include superconducting phases
at the border of magnetism and candidates for genuine non-Fermi
liquid metallic states. However, a major uncertainty in these
studies concerns the possible role of pressure inhomogeneities,
that originate, for instance, in the pressure-transmitting medium
and inhomogeneities of the samples. To settle this issue requires
microscopic measurements of the distribution of lattice constants
across the entire sample volume, which, to the best of our
knowledge, has not been available so far.

The perhaps most prominent and controversial example that
highlights the importance of sample inhomogeneities and pressure
tuning is the heavy-fermion superconductor {\urusi}. The reduction
of entropy at a phase transition at $T_0\approx17.5\,{\rm K}$,
discovered in \urusi\ over twenty years ago, is still not
explained \cite{pal85a,map86a,sch86a}. The associated state in
turn is known as 'hidden order' (HO). The discovery of the HO was
soon followed by the observation of a small antiferromagnetic
moment (SMAF), $m_s\approx 0.01-0.04$~$\mu_{\rm B}$ per U atom
\cite{bro87a} then believed to be an intrinsic property of the HO.
The emergence of large-moment antiferromagnetism (LMAF) of
$m_s\approx 0.4$~$\mu_{\rm B}$ per U atom \cite{ami99a}\ under
pressure consequently prompted intense theoretical efforts to
connect the LMAF with the SMAF and the HO. In particular, models
have been proposed that are based on competing order parameters of
the {\em same} symmetry, i.e., linearly coupled order parameters,
in which the SMAF is intrinsic to the HO
\cite{gor92a,agt94a,sha00a,min05a}. This is contrasted by
proposals for the HO parameter such as incommensurate orbital
currents \cite{cha02a}, multipolar order \cite{kis05a}, or
helicity order \cite{var06a}, where HO and LMAF break {\em
different} symmetries.

The symmetry relationship of HO and LMAF clearly yields the key to
unravelling the nature of the HO state \cite{sha00a,min05a}. While
some neutron scattering studies of the temperature--pressure phase
diagram suggest that the HO--LMAF phase boundary ends in a
critical end point \cite{bou05a}, other studies concluded that it
meets the boundaries of HO and LMAF in a bicritical point
\cite{mot03a,uem05a,has08a,mot08a}. This distinction is crucial,
as a critical end point (bicritical point) implies that HO and
LMAF have the same (different) symmetries, respectively
\cite{min05a}. Moreover, there is a substantial disagreement
w.r.t. the location and shape of the HO--LMAF phase boundary (see
e.g. Ref.\,\cite{ami07a}). This lack of consistency is, finally,
accompanied by considerable variations of the size and pressure
dependence of the moment reported for the SMAF \cite{ami07a},
where NMR studies in powder samples suggested the SMAF to be
parasitic \cite{mat03a}. Accordingly, to identify the HO in
{\urusi} it is essential to clarify unambiguously the nature of
the SMAF and the symmetry relationship of HO and LMAF.

It was long suspected that the conflicting results are due to a
distribution of lattice distortions due to defects.
Notably, uniaxial stress studies showed
that LMAF is stabilized if the $c/a$ ratio $\eta$ of the
tetragonal crystal is reduced by the small amount
$\Delta\eta_c/\eta\approx 5\cdot 10^{-4}$ \cite{yok05a}. Hence,
the parasitic SMAF may in principle result from a distribution of
$\eta$ values across the sample, with its magnitude depending on
sample quality and experimental conditions. In particular,
differences of compressibility of wires, samples supports or
strain gauges that are welded, glued or soldered to the samples
will forcibly generate uncontrolled local strains that strongly
affect any conclusions about the SMAF signal (see, e.g., Refs
\cite{mot03a,uem05a,has08a,mot08a}).

In this Letter we report simultaneous measurements of the
distribution and temperature dependence of the lattice constants,
as well as the antiferromagnetic moment, of a pure single crystal
of {\urusi} utilizing a novel neutron scattering technique called
Larmor diffraction (LD). This allowed to study samples that are
completely free to float in the pressure transmitting medium,
thereby experiencing essentially ideal hydrostatic conditions at
high pressures. Our data of the distribution of lattices constants
$f(\Delta\eta/\eta)$ establishes \textit{quantitatively} that the
SMAF is purely parasitic. In addition, we find a rather abrupt
transition from HO to LMAF which extends from $T\!=\!0$ up to a
bicritical point (preliminary data of $T_N(p)$ were reported in
\cite{nikl09}). We conclude that the HO--LMAF transition is of
first order and that HO and LMAF must be coupled non-linearly.
This settles the perhaps most important, long-standing
experimental issue on the route to identifying the HO.

\pgnfigure{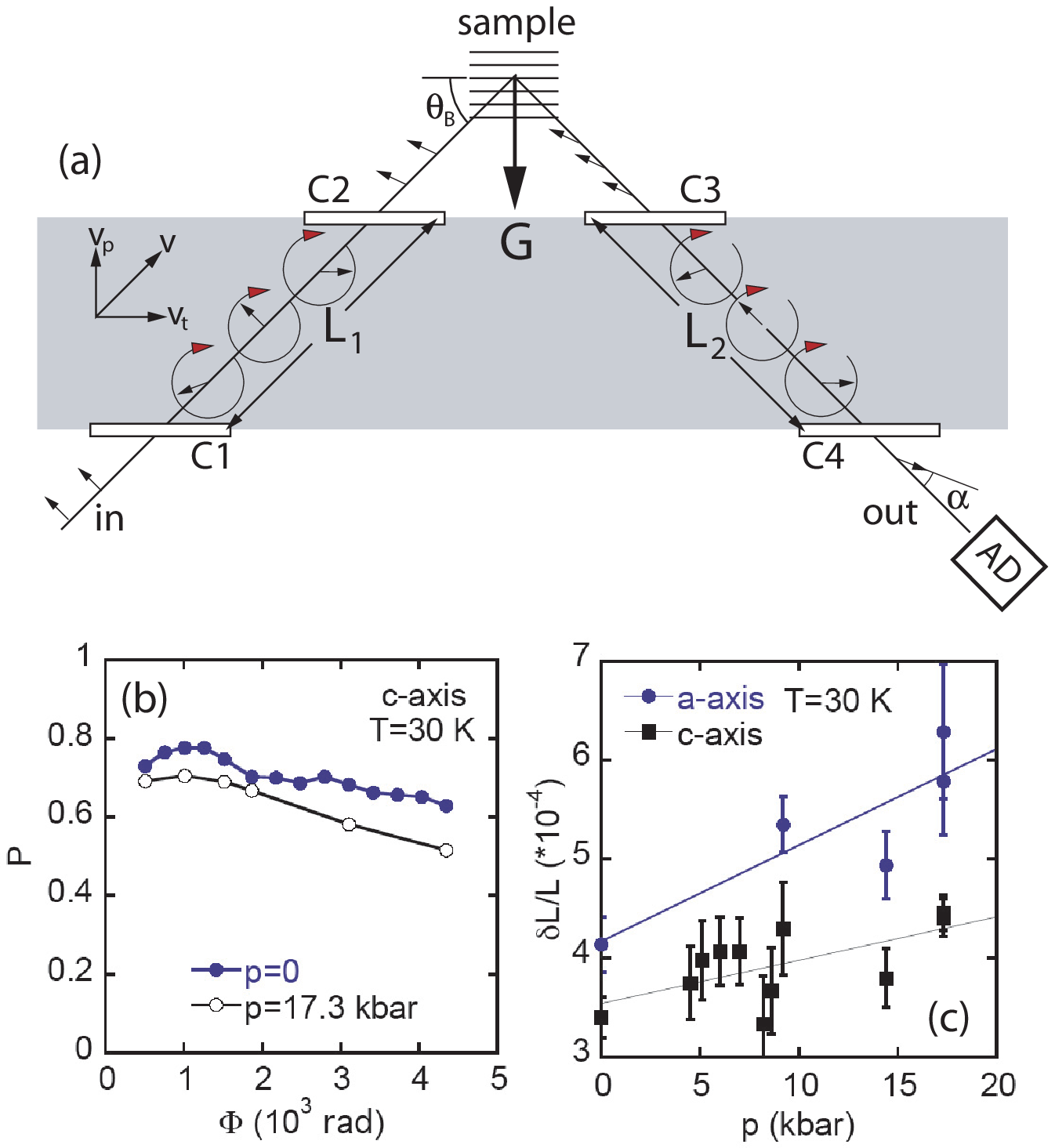}{(a) Schematic of Larmor diffraction
\cite{rek01a,kel02a}, see text for details. (b) Typical variation of the
polarization $P$ as a function of the total Larmor phase $\Phi$.
(c) Pressure dependence of the width of the lattice-constant
distribution for the a- and c-axis in {\urusi}. With increasing
pressure the width of the distribution increases.}

Larmor diffraction permits high-intensity measurements of lattice
constants with an unprecedented high resolution of $\Delta
a/a\approx 10^{-6}$ \cite{rek01a,kel02a}. As shown in
Fig.\,\ref{figure1}\,(a) the sample is thereby illuminated by a
polarized neutron beam (arrows indicate the polarization); $G$ is
the reciprocal lattice vector; $\theta_B$ is the Bragg angle; 'AD'
is the polarization analyzer and detector. The radio frequency
(RF) spin resonance coils (C1-C4) change the polarization
direction, as if the neutrons undergo a Larmor precession with
frequency $\omega_L$ along the distance $L=L_1+L_2$. The total
Larmor phase of precession $\Phi$ thereby depends linearly on the
lattice constant $a$: $\Phi=2\omega_LLma/(\pi\hbar)$ ($m$ is the
mass of the neutron \cite{rek01a,kel02a}).

The Larmor diffraction was carried out at the
spectrometer TRISP at the neutron source FRM\,II. The temperature
and pressure dependence of the lattice constants was inferred from
the (400) Bragg peak for the $a$ axis and the (008) Bragg peak for
the $c$ axis. The magnetic ordered moment was monitored with the
same setup using conventional diffraction.
For our high-pressure studies a Cu:Be clamp cell was used with a
Fluorinert mixture \cite{pfl05b}. The
pressure was inferred at low temperatures
from the (002) reflection of graphite as well as absolute changes
of the lattice constants of {\urusi} taking into account published
values of the compressibility.

The single crystal studied was grown by means of an optical
floating-zone technique at the Amsterdam/Leiden Center. High
sample quality was confirmed via X-ray diffraction and detailed
electron probe microanalysis. Samples cut-off from the ingot
showed good resistance ratios (20 for the $c$ axis and $\approx$
10 for the $a$ axis) and a high superconducting transition
temperature $T_c\approx 1.5$\,K. The magnetization of the large
single crystal agreed very well with data shown in
Ref.\,\cite{pfl06a} and confirmed the absence of ferromagnetic
inclusions. Most importantly, in our neutron scattering
measurements we found an antiferromagnetic moment $m_s \approx
0.012~\mu_B$ per U atom, which matches the smallest moment
reported so far \cite{ami07a}.


As the Larmor phase $\Phi$ is proportional to the lattice constant
$a$, the polarization $P$ of the scattered neutron beam
reflects the distribution of lattice constants
\textit{across the entire sample volume} \cite{kel02a}.
While Larmor diffraction was recently employed for the
first time in a high-pressure study
\cite{pfl07a}, measurements of the distribution of lattice have
not been exploited to resolve a major scientific issue (for proof of
principle studies in Al-alloys see Ref.\,\cite{rek01a}).

In order to explore the origin of the SMAF moment we have measured
the spread of lattice constants, keeping in mind that a
distribution of the $c/a$ lattice-constant ratio
$\eta$ may be responsible for AF order in parts of the sample. As shown in
Fig.\,\ref{figure1}\,(b) $P(\Phi)$ changes
only weakly between ambient pressure and 17.3\,kbar. Accordingly
the distribution of lattice constants, which is given by the
Fourier transform of $P(\Phi)$, changes only weakly as a function
pressure as shown in Fig.\,\ref{figure1}c). Thus pressure only slightly
boosts the distribution of lattice constants, but does not
generate substantial additional inhomogeneities. Moreover, we
confirmed that the size of $m_s$ remained unchanged tiny after our
high-pressure studies.

Assuming a Gaussian distribution of both lattice constants, we
infer the distribution of $\eta$. At low pressure, we arrive at a
full width at half-maximum of $f(\Delta\eta/\eta)_{\rm
FWHM}\approx3.8\cdot 10^{-4}$. Recalling that the tail of the
Gaussian distribution beyond $\Delta\eta_c/\eta\approx 5\cdot
10^{-4}$ represents the sample's volume fraction in which LMAF
forms \cite{yok05a}, an average magnetic moment of
$0.015(5)\mu_{\rm B}$ is expected in our sample. This is in
excellent quantitative agreement with the experimental value and
represents the first main result we report.

\pgnfigure{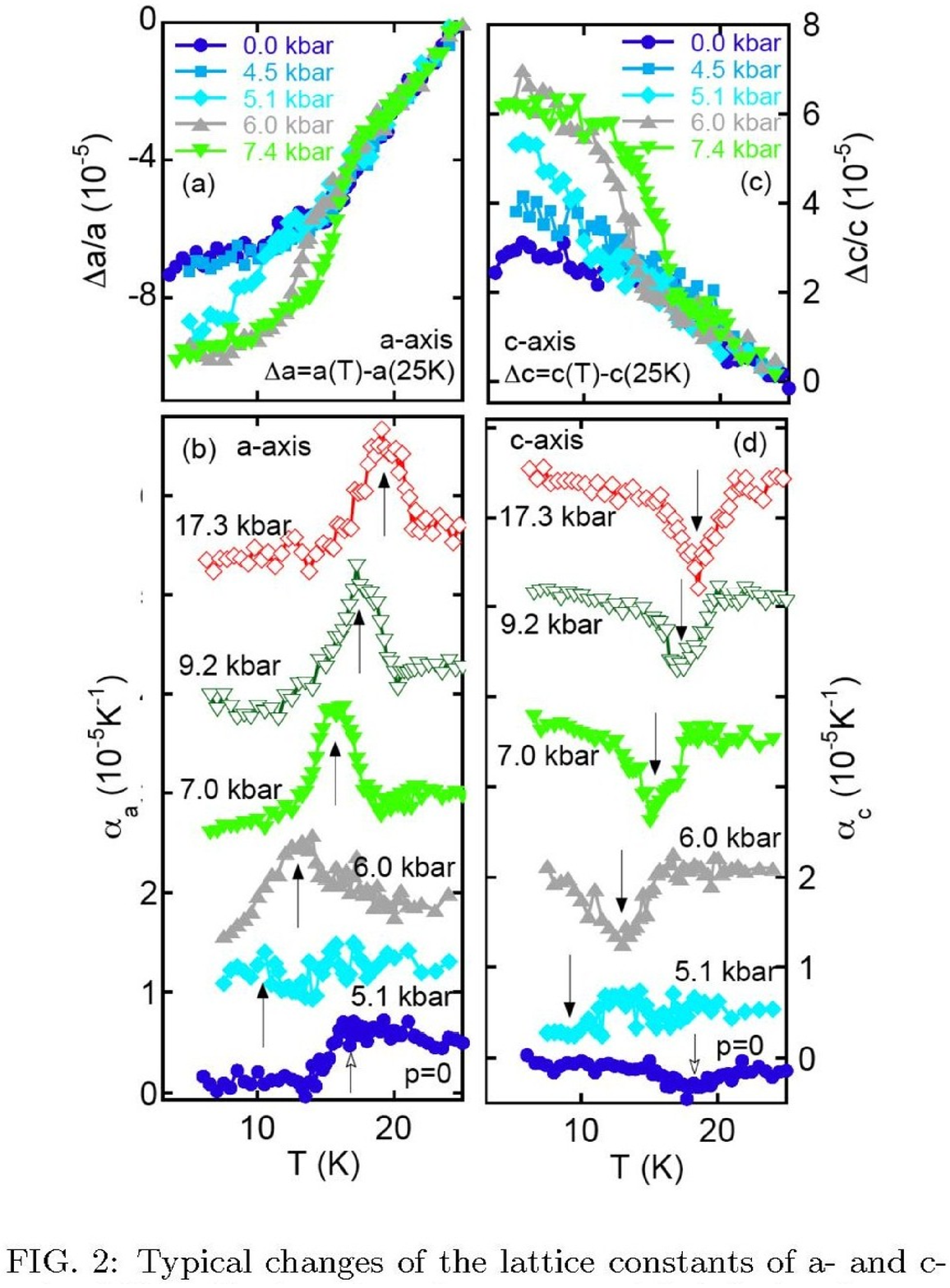}{Temperature dependence of the lattice
constants and thermal expansion at various pressures. The HO and
LMAF transitions are indicated by empty and filled arrows,
respectively. (a) Data for the $a$-axis; (b) thermal expansion of
the $a$-axis derived from the data shown in (a). (c) Data for the
$c$-axis; (d) thermal expansion of the $c$-axis derived from the
data shown in (c).}

We continue with a discussion of the temperature dependence of the
lattice constants and its change with pressure. At zero pressure
in a wide temperature range below room temperature (not shown) the
thermal expansion is positive for both axes \cite{vis86a}.
However, for the c-axis, the lattice constant shows a minimum
close to 40~K and turns negative at lower temperatures. This
general behavior is unchanged under pressure.

Shown in Fig.~\ref{figure2} are typical low-temperature data for
the a- and c-axis. At ambient pressure $T_0$ can be barely
resolved. However, when crossing $p_c\approx4.5\,{\rm kbar}$ a
pronounced additional signature emerges rapidly and merges with
$T_0$. Measurements of the antiferromagnetic moment (see below),
identify this anomaly as the onset of the LMAF order below its
$T_N$. At the transition the lattice constant for the a- and
c-axis show a pronounced contraction and expansion, respectively.

The pressure and temperature dependence of
$m_s$ determined at $(100)$ is shown in the inset of
Fig.\,\ref{figure3}\,(a). Close to $p_c\approx 4.5\,{\rm kbar}$,
where $T_N$ is near base temperature, we observe first evidence of
the LMAF magnetic signal, which rises steeply and already reaches
almost its high-pressure limit at 5~kbar. The pressure dependence
was determined by assuming the widely reported high-pressure value
of $m_s=0.4\mu_{\rm B}$ and comparing the magnetic (100) and
nuclear (004) peak intensity.

\pgnfigure{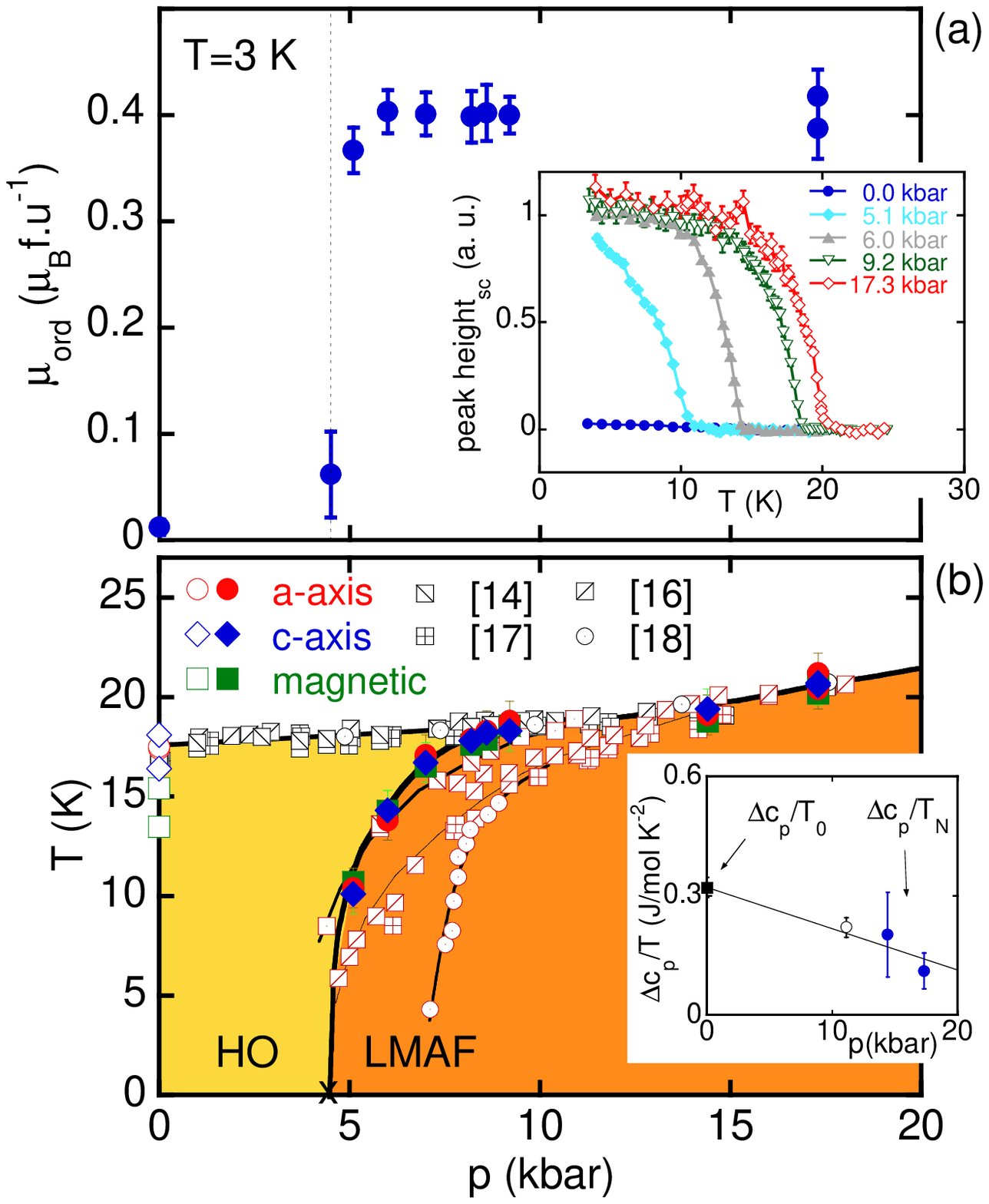}{(a) Pressure dependence of the low-temperature
magnetic moment. The inset shows the temperature dependence of the
magnetic peak height at various pressures. (b) Phase diagram based
on Larmor diffraction and conventional magnetic diffraction data.
The onset of LMAF is marked by full and of HO by empty symbols (x
marks a transition near base temperature). For better comparison
data of $T_N$ (black symbols) from Refs.\,\cite{ami07a,mot03a,has08a,mot08a}
are shown. The $T_0$ values (red symbols) of all
references are consistent. Inset: background-free specific-heat
jump derived from thermal-expansion data via the Ehrenfest
relation (full circles). Heat capacity data is taken from Refs.
\onlinecite{pal85a}\ (square) and \onlinecite{has08a}\ (empty
circle).}

The temperature--pressure phase diagram shown in
Fig.\,\ref{figure3}\,(b) displays $T_0$ and $T_N$ taken from the
magnetic Bragg peak (Fig.\,\ref{figure3}\,(a)) and from the Larmor
diffraction data (Fig.\,\ref{figure2}), respectively. The
different data sets show excellent agreement (except at $p=0$, due
to the variable parasitic nature of the SMAF). The main results
contained in Fig.\,\ref{figure3}\ are (i) the very small value of
0.012~$\mu_B$ of the average low-temperature ordered moment at
zero pressure, (ii) a particularly abrupt increase of the
low-temperature moment at $p_c\approx 4.5$\,kbar (as compared to
previous studies \cite{mot03a,uem05a,has08a,mot08a, ami07a}),
(iii) the steep slope of the HO--LMAF phase boundary, and (iv) the
merging of the HO--LMAF phase boundary with the $T_0$ transition
lines at approximately 9\,kbar. Fig.~\ref{figure2}\,(b) shows that
we can follow this phase boundary from low $T$ up to $T_0$.

\pgnfigure{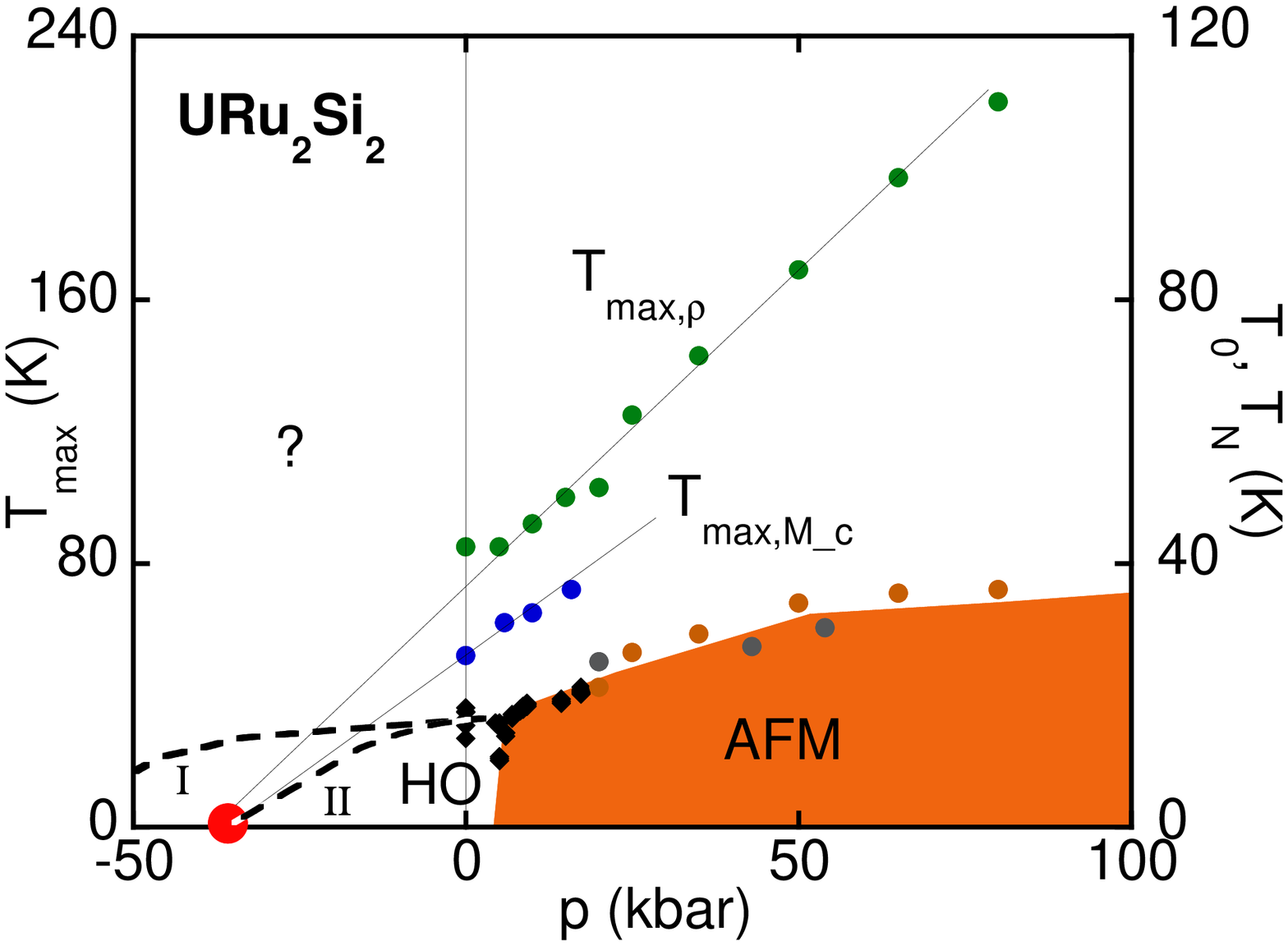}{Extended phase diagram of {\urusi} based on the
data presented here (diamonds) and signatures in resistivity
(brown\cite{kag94a} and grey\cite{has08a}\ circles).
$T_{max,\rho}$ and $T_{max,M_c}$ denote coherence maxima in the
resistivity \cite{kag94a}\ and magnetisation \cite{pfl06a},
respectively. The HO might either mask a QCP (I) or replace LMAF
(II) near quantum criticality.}

Most importantly, (iv) implies that HO and LMAF are fully
separated by a phase boundary which has to be of first order with
a bicritical point, since three second-order phase transition
lines cannot meet in one point. (Note that the phase boundaries
from the HO and LMAF to the disordered high-temperature phase are
already known to be of second order from qualitative heat capacity
measurements \cite{pal85a,has08a}.) This conclusion is perfectly
consistent with (ii) and (iii), and it excludes a linear coupling
between the HO and LMAF order parameters \cite{min05a}.

Although our results show that the HO and LMAF must have different
symmetry, they also suggest that both types of order may have a
common origin. Using the Ehrenfest relation, we have converted our
thermal expansion data into a background-free estimate of the
specific-heat jump, $\Delta C$, at the PM to LMAF transition. The
inset in Fig.\,\ref{figure3}\,(b) shows a continuous evolution
with pressure of the jumps at the LMAF to PM and corresponding HO
to PM transitions. The common origin of both phases may in fact be
related to excitations seen in neutron scattering. Notably,
excitations at (1,0,0) only appear in the HO phase and may be its
salient property \cite{vil08a,elg09a}. However, excitations at
(1.4,0,0) in the PM state become gapped in both the HO and LMAF
state and have been quantitatively linked to the specific heat
jump at zero pressure \cite{wie07a}.

Finally, Fig.\,\ref{figure4} suggests a new route to the HO. It
shows a summary of the pressure evolution of features in the
resistivity and magnetization, which suggest the existence of an
AF quantum critical point (QCP) at an extrapolated negative
pressure. Remarkably, these features, which are usually denoted as
Kondo or coherence temperature, all seem to extrapolate to the
same critical pressure. The HO in \urusi\ could be understood as
emerging from quantum criticality, possibly even masking an AF QCP
(case I in Fig.\,\ref{figure4}). However, considering the
remoteness of the proposed QCP, the balance between LMAF and HO
might rather be tipped by pressure induced changes of the
properties of \urusi. This has for example been suggested in a
recent proposal, where HO and LMAF are believed to be variants of
the same underlying complex order parameter \cite{hau09a}. In such
a scenario, the HO is more likely to collapse at the proposed QCP
as denoted by the dashed line (case II in Fig.\,\ref{figure4}).
The pressure dependence of $\Delta C$ sets constraints on any
theory of the HO involving quantum criticality.

In conclusion, using Larmor diffraction to determine the lattice
constants and their distribution at high pressures and with very
high precision, we were able to show that the SMAF in {\urusi} is
purely parasitic. Moreover, the HO and LMAF are separated by a
line of first-order transitions ending in a bicritical point. This
is the behavior expected for the HO being, e.g., incommensurate
orbital currents, helicity order, or multipolar order.

We are grateful to P. B\"oni, A. Rosch, A. de Visser, K. Buchner,
F. M. Grosche and G. G. Lonzarich for support and stimulating
discussions. We thank FRMII for general support. CP and MV
acknowledge support through DFG FOR 960 (Quantum phase
transitions) and MV also acknowledges support through DFG SFB 608.

\end{document}